\documentclass[prd,aps,twocolumn,superscriptaddress,showpacs,floatfix,amsmath,amssymb]{revtex4}

\newcommand{\be}{\begin{eqnarray}}
\newcommand{\ee}{\end{eqnarray}}
\newcommand{\eq}[1]{(\ref{#1})}

\newcommand{\cF}{{\cal F}}
\newcommand{\cB}{{\cal B}}
\newcommand{\cG}{{\cal G}}

\newcommand{\sC}{{\mathsf{C}}}
\newcommand{\cY}{{\mathcal{Y}}}

\newcommand{\sS}{{\mathsf S}}
\newcommand{\cL}{{\cal L}}

\newcommand{\vect}[2]{\left(\begin{array}{c} #1 \\[1mm] #2 \end{array}\right)}

\begin{document}

\title{Baryon number violation and a new electroweak interaction}

\author{M. N. Chernodub}
\affiliation{LMPT,
CNRS UMR 6083, F\'ed\'eration Denis Poisson, Universit\'e de Tours,
Parc de Grandmont, F37200, Tours, France}
\affiliation{Department  of Mathematical Physics and Astronomy,
Ghent University, Krijgslaan 281, S9, B-9000 Ghent, Belgium}
\affiliation{ITEP,
B.~Cheremushkinskaya 25, Moscow, 117218, Russia}
\author{Antti J. Niemi}
\affiliation{LMPT,
CNRS UMR 6083, F\'ed\'eration Denis Poisson, Universit\'e de Tours,
Parc de Grandmont, F37200, Tours, France}
\affiliation{Department of Theoretical Physics, Uppsala
University, P.O. Box 803, S-75108, Uppsala, Sweden}
\affiliation{Chern Institute of Mathematics,
Nankai University, Tianjin 300071, People's Republic of China}

\begin{abstract}
We introduce a new supercurrent in the electroweak sector of the
standard model. Its interaction with the hypergauge field
influences the mass of the $Z$ boson but has no
effect on the $W^\pm$-boson masses. In the
leptonic sector it affects the numerical value of the vector
and axial coupling constants between neutral currents and
the $Z$ boson, and a comparison with experimental values
yields an upper bound to the strength of the coupling between
the supercurrent and the hypergauge field.
In the baryonic sector the supercurrent gives a new contribution to the anomaly
equation for baryon number current. As a consequence it may have
an effect on baryogenesis.
\end{abstract}

\pacs{11.15.Ex,12.15.-y,11.27.+d,98.80.Cq}

\date{\today}

\maketitle

In the conventional superconductor the supercurrent $\vec J$
is the gauge-invariant vector field~\cite{ref:volume9},
\begin{equation}
\vec J \ = \ e \psi^\star
(  i \frac{\hbar }{2m} \stackrel{\leftrightarrow}{\nabla}
- \frac{2e}{mc} \vec A ) \psi
\ \equiv \ e |\psi|^2 \vec C\,,
\label{suco}
\end{equation}
where
$\vec A$ is the Maxwellian $U(1)$ gauge field and $\psi$ is the wave function of
Cooper pairs. When we use this relation to eliminate $\vec A$
in favor of $\vec J$ in the Landau-Ginzburg Hamiltonian that describes the conventional
superconductivity
\[
\mathcal H = \frac{1}{2} {\vec B}^2 + |(\vec \nabla - 2ie \vec A) \psi |^2 + \lambda ( | \psi |^2
- v^2)^2
\]
 where $\vec B$ is the magnetic field, we obtain a Hamiltonian that involves only the vector
field $\vec C$, the Cooper pair density $\rho = |\psi|$ and the phase of the Cooper pair
$\theta = \arg \psi$ as independent
and manifestly gauge-invariant field variables,
\begin{equation}
\mathcal H = \frac{1}{4} \left(  C_{ij} + \frac{\pi}{e} \tilde \sigma_{ij}\right)^2  + (\vec \nabla \rho)^2 + \rho^2 {\vec C}^2 + \lambda
(\rho^2 - v^2)^2
\label{elim1}
\end{equation}
with
\[
C_{ij} = \nabla_i C_j - \nabla_j C_j
\]
and
\begin{equation}
\tilde \sigma_{ij} = \epsilon_{ijk} \sigma_k = \frac{1}{2\pi} [\nabla_i , \nabla_j] \theta
\label{dirst}
\end{equation}
is the string current with support that coincides with the world sheet of the core of
the Abrikosov vortex. When (\ref{elim1}) describes such a vortex, (\ref{dirst}) subtracts a similar
singular contribution that appears in $C_{ij}$. In the third term this singularity becomes suppressed
since the density $\rho$ vanishes at the core of the vortex.
Furthermore, whenever
the ground state value of $\rho$ is nonvanishing, the vector field $\vec C$ is
massive and becomes subject to the Mei\ss ner effect.

In the present paper we are interested in the physical consequences when a non-Abelian
generalization of (\ref{suco}), (\ref{elim1}) is implemented in the Weinberg-Salam model of electroweak
interactions.
Indeed, the structure of the Weinberg-Salam Lagrangian
closely resembles that of the conventional superconductor: Now the gauge field and the
Higgs field both transform under the non-Abelian gauge group
$G^{}_{EW} = SU^{}_L(2) \times U^{}_Y(1)$, and the Higgs field supports three
independent angular (phase) variables that can be combined with
three of the gauge fields into the neutral
$Z$ boson and the charged $W^\pm$ bosons. In
the low temperature phase  these vector fields become massive, in parallel with the
Mei\ss ner effect of superconductivity.

In the sequel our motivation lies
in the following: In the case of an ordinary superconductor,
the supercurrent can be used to entirely remove the gauge field.
But in the case of the electroweak theory there are {\it four} independent
gauge fields. As a consequence there may be as many as four independent
supercurrents~\cite{ref:faddeev}.
We shall be mainly interested in the properties of a
scrupulously chosen supercurrent that appears to lead to
a new electroweak interactions with topological ramifications. In
particular, we propose that our supercurrent affects the anomaly equation for the
baryon number current in a manner that enhances baryon production.
This effect is catalyzed by the presence of the (embedded) topological
defects, the Nambu monopoles, in the electroweak model~\cite{ref:Nambu}.

Presently, there are many approaches for explaining baryon
asymmetry of the Universe~\cite{ref:Baryon}, \cite{ref:MSSM}.
According to the hot baryogenesis
scenario~\cite{ref:hot:Baryogenesis}, the cooling
of the early Universe produced baryons by
sphaleron driven thermal activation~\cite{ref:sphaleron}.
But the Standard Model predicts that instead
of a finite temperature phase transition, the cooling of the early Universe
proceeded through a crossover transition that retained
a thermodynamic equilibrium and thus the hot baryogenesis
contrasts the third Sakharov
condition~\cite{ref:Sakharov}. The alternative, cold
baryogenesis scenario~\cite{ref:tachyonic:theory,ref:tachyonic:exp}
allows for fluctuations that may have caused the early Universe
to abandon a state of thermal equilibrium. This may have occurred
during the inflationary epoch of the early
Universe, at a TeV scale and in combination with the tachyonic
transition~\cite{ref:HybridInflation}. Indeed, if the inflaton couples
to the Higgs field in a manner that forces the effective Higgs
mass to change its sign, a sudden and rapid production
of a large amount of particles could take place.
Tachyonic preheating~\cite{ref:preheating} may thus
lead to a net baryon number production due to a change in the
Chern-Simons (CS) contribution to the anomaly equation for the baryon number current.

In both of these scenarios a central role is played
by various embedded defects or their bound states.
For example the electroweak sphaleron that drives
hot baryogenesis can be interpreted as a pair
of a Nambu monopole \cite{ref:Nambu} and an\-ti\-monopole, bound to
each other by a $Z$-vortex~\cite{ref:connected}. It is also presumed
that both Nambu monopoles and electroweak vortices were copious
during the hot phase~\cite{ref:hot:phase:objects}.
Additional defects such as textures~\cite{ref:tachyonic:theory},
half-knots~\cite{ref:tachyonic:exp}, knotted hypermagnetic
fields~\cite{ref:HK}, linked and twisted
vortices~\cite{Vachaspati:1994ng}, center vortices~\cite{ref:Jeff} {\it etc.}
may also have relevance.

We start by recalling how the electroweak gauge group
$G_{\mathrm{EW}}$ acts on the non-Abelian gauge field $\hat A_\mu
\equiv {\vec A}_\mu \cdot \vec \tau$,
the hypergauge field $Y_\mu$ and the Higgs field $\Phi$,
\begin{equation}
G_{\mathrm{EW}}: \quad \left\{ {
\begin{array}{lcl}
{\hat A}_\mu
& \to & \Omega {\hat A}_\mu \Omega^\dagger -
\frac{2 i}{g} \Omega \partial_\mu \Omega^\dagger \\[1mm]
Y_\mu & \to & Y_\mu - \frac{2}{g'} \partial_\mu \omega_Y \\
\Phi & \to & \exp\{i \omega_Y\}\, \Omega\,\Phi
\end{array}
} \right.
\label{trans}
\end{equation}
Here $\Omega$ is the $SU_L(2)$ gauge matrix,
$\omega_Y$ parameterizes a non-compact $U_Y(1)$
hypergauge transformation, and $g$ and $g'$ are
the two couplings of the Weinberg-Salam model with
the Weinberg angle given by $\tan \theta_W = g'/g$.
We now employ these transformation laws to introduce an electroweak
supercurrent: The hypercharge transformation
has no effect on the $SU_L(2)$ gauge field ${\hat A}_\mu$, while the Higgs
field has a nontrivial hypercharge. Thus we can use
these fields to construct a supercurrent in a direct
non-Abelian generalization of  (\ref{suco})
\cite{foot},
\be
\mathcal J_\mu =
\frac{2}{g'} \Phi^\dagger
\Bigl(- \frac{g}{2} {\vec A}_\mu \cdot {{\vec\tau}}
+ i \partial_\mu \Bigr)\Phi = \Phi^\dagger \Phi \cdot \cY_\mu
\label{n1}
\ee
We also introduce the unit vector
\be
\vec n \ =
\ - \frac{\Phi^\dagger \vec \tau \Phi}{\Phi^\dagger \Phi}\,,
\label{n}
\ee
that determines the direction of the isospin
polarization in the $SU_L(2)$ gauge group.

It is important to notice that the non-Abelian
supercurrent $\cY_\mu$ in (\ref{n1}) is $SU_L(2)$ gauge-invariant but under $U_Y(1)$ it transforms like the hypergauge field~$Y_\mu$ so that
\be
U_Y(1): \qquad
\cY_\mu & \to & \cY_\mu - \frac{2}{g'} \partial_\mu \omega_Y\,.
\label{eq:trans:hyper}
\ee

Thus we can modify the Lagrangian of the standard electroweak model by
executing the following {\it shift} of $Y_\mu$
\be
Y_\mu \to (1 - \kappa) Y_\mu + \kappa \cY_\mu\,,
\label{eq:shift}
\ee
in the Lagrangian. In Eq.~\eq{eq:shift} $\kappa$ is an {\it ab initio} free parameter.
This leads to a modification of the Lagrangian by a
gauge-invariant operator.
Indeed, since the gauge transformation properties of the
original and shifted $Y_\mu$
are identical, the shift (\ref{eq:shift}) preserves all local gauge symmetries
of the original electroweak Lagrangian.

Furthermore, this shift~\eq{eq:shift} can be introduced independently in the Higgs sector
and in the fermionic sector of the Lagrangian ($\kappa_\Phi \not= \kappa_f$),
and the parameter $\kappa$ may also be different for different fermionic
flavors ($\kappa_f \not= \kappa_{f'}$).

We shall now consider the consequences of (\ref{eq:shift}) in the
ensuing {\it low energy effective} Weinberg-Salam model.
In particular, we shall argue that in the fermionic sector
of the electroweak Lagrangian, the shift (\ref{eq:shift}) may have a
profound effect on the anomaly equation that governs the
baryon number nonconservation in early universe baryogenesis. But we
first inspect whether (\ref{eq:shift}) leads to other observable
effects, and, in particular, whether we can obtain an estimate for the numerical
value of the parameter $\kappa$.

The kinetic term of the hypergauge field $\cY_\mu$ is a
gauge-invariant quantity and thus it may be independently added to the
Lagrangian. But here we restrict our attention solely to the effects
of the shift (\ref{eq:shift}) to the interactions between the Higgs,
vector boson, and fermion fields.

We first observe that if we introduce the shift~\eq{eq:shift} (with parameter $\kappa_\Phi$)
in the Higgs kinetic term, this leads to a correction of the $Z$-boson mass,
\be
M^2_Z = \left[\frac{g v (1 - \kappa_\Phi)}{2 \cos \theta_W}\right]^2\,,
\label{eq:Z1:shift}
\ee
and has no effect to the $W^\pm$-boson mass.

We proceed to the fermionic sector of the
electroweak Lagrangian. There are  $N_f$ flavors of
the left-handed and right-handed fermions, and
in the following we shall not always differ between leptons and quarks. If we
implement the shift
(\ref{eq:shift}) (with $\kappa_f \equiv \kappa$), besides rescaling the
familiar hypergauge interaction by a factor $(1-\kappa)$  it leads to a new
interaction between the Higgs field, the $SU_L(2)$ gauge vectors
and the fermions which is of the form
\be
\frac{i}{2} g' \kappa \, {\mathsf Y}_\ell  {\bar \Psi}_{f,\ell} \gamma^\mu
\cY_\mu \,\Psi_{f,\ell}\,.
\label{eq:L:cY}
\ee
Here $\ell = L,R$, and the lepton hypercharges are ${\mathsf Y}_L=-1$
and ${\mathsf Y}_R=-2$.
The essential feature of (\ref{eq:L:cY}) is that it
leaves the interactions between the charged weak currents and the
$W^\pm$ bosons unchanged. But the neutral sector of the electroweak
Lagrangian
\be
\cL^{(0)}_\psi = - g J^\mu_3 W^3_\mu - (g'/2)\, J^\mu_Y \, Y_\mu
\ee
becomes modified by the
shifts in the neutral currents
\be
J^\mu_3 \to J^\mu_3 + \kappa\,J^\mu_Y/2 \quad \mathrm{and}
\quad J^\mu_Y \to (1-\kappa)J^\mu_Y\,.
\ee
When we specify $f=e,\nu$, then
\[
\begin{array}{rcl}
J^\mu_3 & = & \hspace{3mm} \bigl(\bar\psi_{\nu,L}\gamma^\mu\psi_{\nu,L}
- \bar\psi_{e,L}\gamma^\mu\psi_{e,L}\bigr)/2\,,\\[1mm]
J^\mu_Y & = & - \bigl(\bar\psi_{\nu,L}\gamma^\mu\psi_{\nu,L}
+ \bar\psi_{e,L}\gamma^\mu\psi_{e,L} +
2 \bar\psi_{e,R}\gamma^\mu\psi_{e,R}\bigr)\,,
\end{array}
\]
and we find that the coupling between the electric current
and the electromagnetic gauge field still leads to the standard definition
of the Weinberg angle $\theta_W$, with electric charge
given by $e = g \sin \theta_W$.
Defining the field combinations in the usual manner,
\be
\vect{A_\mu}{Z_\mu} =
\left(
\begin{array}{cc}
\cos\theta_W & \sin\theta_W\\
- \sin\theta_W & \cos\theta_W
\end{array}
\right)
\vect{Y_\mu}{A^3_\mu}\,,
\label{eq:rotation}
\ee
we then get a familiar form for the Lagrangian that describes
the neutral current sector,
\[
\cL^{(0)}_L \! = \! j^\mu_e A_\mu
- \frac{g}{2 \cos \theta_W} \sum_{f=e,\nu} \bar \Psi_f
\gamma^\mu (g^{(f)}_V - g^{(f)}_A \gamma^5) \Psi_f \, Z_\mu\,.
\]
But now the vector $g_V$ and axial $g_A$ couplings of the neutral currents
to the $Z$-boson field have been modified,
\be
\begin{array}{rcl}
g^{(f)}_V & = & (1 - \kappa) \, T^{(f)}_{3}
- 2 Q^{(f)} (\sin^2 \theta_W - \kappa_f)\,,\\[1mm]
g^{(f)}_A & = & (1 - \kappa) \, T^{(f)}_{3}\,.
\end{array}
\ee
Here $T^{(f)}_{3}$
is the weak isospin ($T^{(\nu)}_3 = +1/2$ for
$f = \nu$ and $T^{(\ell)}_3 = -1/2$ for $f = \ell \equiv e,\mu,\tau$), and $Q^{(f)}$
is the electric charge ($Q^{(\nu)} = 0$ and $Q^{(\ell)} = -1$).
When we use the uncertainties in constraints on the
$g_{V,A}$ couplings published by the Particle Data Group~\cite{ref:PDG}
we get an estimate for the experimentally allowed value of $\kappa$,
\be
|\kappa^{(\ell)}| \lesssim 3 \times 10^{-4},
\ \
|\kappa^{\nu_e}| \lesssim 10^{-1},
\ \
|\kappa^{\nu_\mu}| \lesssim 2\times 10^{-2}.
\ee

In the leptonic sector the shift~\eq{eq:shift} has no effect
on the electromagnetic interaction involving the massless gauge field $A_\mu$.
There are also no changes in the charged weak interactions that involve the
charged currents $W^\pm_\nu$.  In particular, there is no effect
on the fermion masses: The only effect of the shift~\eq{eq:shift} appears in
the neutral weak sector involving the exchange of the $Z$ boson.

However, when we proceed to the baryonic sector we find that
the shift~\eq{eq:shift}
does have an impact on the processes that violate $B+L$
conservation. Thus it may well have an influence on
baryogenesis. For this
we recall the standard nonconservation (anomaly) equation
for the baryon number current
which in terms of our shifted field becomes
\be
\partial_\mu j^\mu_B & = & \frac{N_f}{32 \pi^2} \,\Bigl(- g^2
{\vec G}_{\mu\nu} {\widetilde {\vec G}}_{\mu\nu}
+ {g'}^2 F^{(\kappa)}_{\mu\nu} {\widetilde F}^{(\kappa)}_{\mu\nu}\Bigr)\,.
\label{eq:anomaly}
\ee
Here
\[
F^{(\kappa)}_{\mu\nu} = (1 - \kappa) (\partial_\mu Y_\nu - \partial_\nu
Y_\mu) + \kappa \cF_{\mu\nu}
\]
\be
\cF_{\mu\nu} = \partial_\mu \cY_\nu - \partial_\nu \cY_\mu
= \frac{g}{g'} \cG_{\mu\nu}(\vec W, \vec n)
+ \frac{4 \pi}{g'} {\widetilde\Sigma}^{\sS_0}_{\mu\nu}\,.
\label{eq:cF}
\ee
and we have introduced
the gauge-invariant 't~Hooft tensor~\cite{ref:thooft},
\be
\cG_{\mu\nu}(\vec A, \vec n) = \vec n
\cdot \vec{G}_{\mu\nu} - \frac{1}{g} \vec n \cdot
{\mathrm D}^{}_\mu  \vec n \times
{\mathrm D}^{}_\nu \vec n\,,
\ee
where $\vec n$ is defined in (\ref{n}),
\[
\vec{G}_{\mu\nu} = \partial_\mu \vec A_\nu - \partial_\nu \vec A_\mu
+ g \vec A_\mu \times \vec A_\nu
\]
and ${\mathrm D}^{ab}_\mu = \delta^{ab}\partial_\mu + g \epsilon^{acb}A^c_\mu$.
The tensor $\Sigma^{\sS_0}_{\mu\nu}$ denotes a (Dirac-like)
string contribution that describes the two dimensional
worldsheet of a closed $Z$--vortex~\cite{ref:Z:vortex,ref:embedded:defects}.
In the unitary gauge
where $\vec n$ becomes aligned with the (positive) $z$ axis
in the isospin space, the 't~Hooft tensor is
\be
\cG_{\mu\nu} = \partial_\mu W^3_\nu - \partial_\nu W^3_\mu + \frac{4 \pi}{g}
{\widetilde \Sigma}^{\sS'_\sC}_{\mu\nu}\,,
\label{eq:cG:Unitary}
\ee
where $\sS'_\sC$ denotes the conventional Dirac string of the Nambu
monopole. As a consequence
the total world surface determined by the string structures
of the Nambu monopoles
is $\sS_\sC = \sS_0 + \sS'_\sC$. In this gauge we can write
\be
F^{(\kappa)}_{\mu\nu} & = & \partial_{\mu} \cB_{\nu} - \partial_\nu \cB_{\mu}
+ \frac{4 \pi \,\kappa}{g'} \,
{\widetilde \Sigma}^{\sS_\sC}_{\mu\nu} \,,\\
\cB_\mu & = & \cos\theta_W \, A_\mu -
\Bigl(\sin\theta_W - \frac{\kappa}{\sin\theta_W}\Bigr)\, Z_\mu\,.
\label{eq:cB}
\ee
Here $\cB_\mu$ is an Abelian gauge field without Dirac singularities.
Then we introduce the conserved ($\partial^\mu k_\mu^{\sC} = 0$) current
of the Nambu monopole,
\[
k_\mu^{\sC} \ = \ \kappa \cdot
\frac{2\pi }{g'}
{\epsilon_{\mu}}^{\nu\rho\sigma} \partial_\nu
{\widetilde \Sigma}^{\sS_\sC}_{\rho\sigma}.
\]

When we integrate both sides of~\eq{eq:anomaly} over the four-dimensional
volume between two distant, time separated three-dimensional
spatial surfaces, we obtain an equation
for the net production of baryon number,
\be
\Delta Q_B = N_f \left(\Delta N_{\mathrm{CS}} -
\gamma_\kappa
\Delta n_{\mathrm{CS}}
+ \frac{e \kappa}{4 \pi} {\mathsf \Phi}^\sC
+ \frac{\kappa^2}{2} N_{\sS_\sC}
\right)
\label{eq:delta:Q}
\label{Q}
\ee
The first two terms are the conventional results [modulo the
$\gamma_\kappa \equiv (1-\kappa)^2$ factor],
reflecting the change in
the non-Abelian Chern-Simons number $N_{\mathrm{CS}}$
and the Abelian Chern-Simons number $n_{CS}$.
These terms also contain contributions from linked
and twisted $Z$-vortices~\cite{Vachaspati:1994ng}.

The third and fourth terms in Eq.~\eq{Q} are new and entirely
due to our supercurrent shift. The third term has its origin in the
presence of Nambu monopoles. Using the definition of the monopole current
and assuming that $\sC$ is a large enough trajectory so that the contribution
from the massive $Z$-boson field can be ignored we get
\be
{\mathsf \Phi}^\sC \ = \ \frac{1}{\cos\theta_W} \int {\mathrm d}^4 x\,
k^\sC_\mu \cB^\mu = \oint_\sC {\mathrm d}x_\mu A^\mu + \dots\,.
\label{eq:flux}
\ee
Thus ${\mathsf \Phi}^\sC$ coincides with the standard electromagnetic flux
that emanates from the string structure of the Nambu monopoles.
The third contribution to the baryon number production (\ref{Q}) can formally
be interpreted as the Witten effect~\cite{ref:Witten}, as it
makes the Nambu monopoles to serve as dyons with electric charge $q_e \propto e \kappa$.

The fourth term in (\ref{Q}) counts the
number of (self-) intersections of the string
worldsheets~\cite{ref:Self:intersections}:
\[
N_I[\sS_\sC] = \frac{1}{2} \int {\mathrm d}^4 x \,
\Sigma^{\sS_\sC}_{\mu\nu} {\widetilde \Sigma}^{\sS_\sC}_{\mu\nu}\,.
\]
This quantity  receives contributions both from
transversal intersection points
and from the twisting points, and it can be entirely
presented in terms of the evolution of
the writhing number of the surface~\cite{ref:Self:intersections}.
Note that while the ensuing contribution to the change
in the baryon number depends on $\kappa$, it is entirely
independent of the couplings $g$ and $g'$ of the
electroweak Lagrangian.

We conclude with a few remarks:
A shift of the form (\ref{eq:shift}) introduces a modification
of the electroweak theory by a gauge-invariant operator that
appears to be fully consistent with the local symmetries
of the Weinberg-Salam Lagrangian.
The energy scale of the new transition may be different from the
electroweak scale in which case the $Z$ boson might either remain massive
above the electroweak transition, or acquire the additional
mass correction~\eq{eq:Z1:shift} at some lower energy scale.

Note that in the leptonic sector the sole effect of the shift appears to be in
a correction to the value of the vector and axial couplings of
neutral currents - this is a tree-level effect.

Finally, in the baryonic sector of the electroweak theory the shift
(\ref{eq:shift}) appears to
affect the anomaly in
the baryon current conservation. In particular, if the early
Universe went through a period with copious production and subsequent
annihilation of Nambu monopoles, the
shift could have played a central r\^ole in baryogenesis.
Being of a non-perturbative nature, this aspect deserves to be studied
in greater detail numerically within an adopted model of electroweak
history of the Universe.

This work has been supported by a STINT Institutional Grant No. IG2004-2 025.
The work by M.N.C. is also supported by Grants No. RFBR 05-02-16206a, No.
RFBR-DFG 06-02-04010, and by a CNRS grant. The work by A.J.N. is also
supported by a VR Grant No. 2006-3376 and by the Project Grant ANR NT05-142856.
The authors thank L.D. Faddeev for discussions.


\begin{thebibliography}{99}
\bibitem{foot}
Following Eq. (\ref{suco}) we normalize the supercurrent by the
Higgs field. In full parallel with the Ginzburg-Landau model,
the apparent nonanalyticity disappears at the level of the Lagrangian.
\end{thebibliography}
\end{document}